\documentclass[twocolumn,showpacs,amssymb,nobibnotes,aps,prl]{revtex4-1}

\usepackage{enumerate,amsmath}
\usepackage{graphicx}
\begin{document}

\title{Anti-Zeno Effect for Quantum Transport in Disordered Systems}
\author{Keisuke Fujii and Katsuji Yamamoto}
\affiliation{
Department of Nuclear Engineering, Kyoto University, Kyoto 606-8501, Japan}

\date{\today}
\begin{abstract}
We demonstrate that repeated measurements in disordered systems
can induce a quantum anti-Zeno effect under certain conditions
to enhance quantum transport.  The enhancement of energy transfer
is really exhibited in multisite models under repeated measurements.
The optimal measurement interval for the anti-Zeno effect
and the maximal efficiency of energy transfer
are specified in terms of the relevant physical parameters.
Since the environment acts as frequent measurements on the system,
the decoherence-induced energy transfer,
which has been discussed recently for photosynthetic complexes,
may be interpreted in terms of the anti-Zeno effect.
We further find an interesting phenomenon in a specific three-site case,
where local decoherence or repeated measurements
may even promote entanglement generation between the nonlocal sites.
\end{abstract}

\maketitle

%%%%%
{\it Introduction.}---
%%%%%
Recently a lot of efforts have been devoted
to obtaining better understanding of quantum systems,
especially from viewpoints of information processing and communication.
A variety of quantum-theoretical analyses for such systems
have been applied in wide range of natural sciences,
extending to biological systems
\cite{Adolphs06,Rebentrost09,Plenio08,Sarovar09,Scholak09,
Cai09,Kominis08,Gauger09},
and the results are supported by recent experimental progress
\cite{Engel07,Lee07,Maeda08}.
An interesting example of quantum effects is found in biological systems,
namely, the highly efficient energy transfer
in light harvesting photosynthetic complexes.
It has been confirmed by experiments
that remarkably long-lived electronic quantum coherence
plays an important role in the energy-transfer process
within the Fenna-Matthews-Olson (FMO) bacteriochlorophyll complex
\cite{Engel07}.
Theoretical studies of the energy transfer
in light-harvesting complexes have also been carried out
\cite{Adolphs06,Rebentrost09,Plenio08,Sarovar09,Scholak09}.
The excitation energies of antenna pigments
in photosynthetic complexes are disordered
due to their interaction with local environments.
In such disordered systems, coherent excitation transfer
is usually suppressed due to destructive interference,
and the population of excitation is localized,
which is well known as the Anderson localization \cite{Localization}.
Thus one might expect naively that the efficiency of energy transfer
is deteriorated crucially in disordered systems.
It has been, however, revealed by recent studies
that quantum coherence and environment-induced decoherence
may collaborate under certain circumstances,
enhancing the energy transfer \cite{Rebentrost09,Plenio08}.

In quantum mechanics, the nature of measurement processes
has been one of the most fundamental problems.
It is a unique feature in quantum mechanics
that measurement disturbs the observed system inevitably.
A familiar example is the quantum Zeno effect,
which suggests that frequent measurements freeze
the time evolution of the system
\cite{Misra-Sudarshan77,Cook88,Itano90,Koshino05}.
Contrarily, it is known as the anti-Zeno effect
that repeated measurements under certain circumstances
may even accelerate physical processes
including decay processes \cite{AntiZeno},
oscillation in two-level systems \cite{Luis03},
and delocalization in disordered systems \cite{Delocalization}.
This acceleration can be understood as a violation of energy conservation
(broadening) due to the consecutive measurements.

In this article, we investigate how repeated measurements affect
quantum transport in disordered systems,
especially from the viewpoint of the quantum (anti-)Zeno effect.
Specifically, we demonstrate that the energy transfer in disordered systems
can be enhanced via the anti-Zeno effect.
The enhancement of energy transfer is really exhibited
in multisite models under repeated measurements.
In particular, the optimal measurement interval for the anti-Zeno effect
and the maximal efficiency of energy transfer
are specified in terms of the relevant physical parameters.
Since the environment acts as frequent measurements on the system,
the decoherence-induced energy transfer, which has been discussed recently
for photosynthetic complexes \cite{Rebentrost09,Plenio08},
may be interpreted in terms of the anti-Zeno effect.
We further find an interesting phenomenon in a specific three-site case,
where local decoherence or repeated measurements
may even promote entanglement generation between the non-local sites.

\begin{figure}
\centering
\scalebox{0.25}
{\includegraphics*[0cm,0.5cm][27cm,10cm]{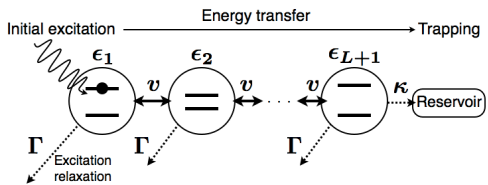}}
\caption{$L+1$ qubits coupled with the nearest neighbors.
The initial excitation at site 1 is transferred
to site $L+1$ through the coupling with strength $v$
and then trapped in the reservoir with rate $\kappa$.
The excitation also decays at each site with rate $\Gamma$.}
\label{model}
\end{figure}

%%%%%
{\it Anti-Zeno effect in disordered systems.}---
%%%%%
We consider a transport problem in a disordered system
consisting of $L+1$ qubits which are coupled specifically
in one dimension with the nearest neighbors (Fig. \ref{model}).
For simplicity, we restrict our attention to the single excitation subspace
spanned by the states $ |i\rangle = |g\rangle^{\otimes i-1}
\otimes |e\rangle \otimes |g\rangle^{\otimes L+1-i} $,
where site $i$ is only excited ($|e\rangle$)
while the other sites are in the ground states ($|g\rangle$).
Then, the Hamiltonian (so called tight-binding Anderson model)
is given by
%%%
\begin{eqnarray}
H = \sum_{i=1}^{L+1} \epsilon_i |i \rangle \langle i|
+ \sum_i^L v (|i \rangle \langle i+1| + |i+1 \rangle \langle i|) ,
\label{Anderson}
\end{eqnarray}
%%%
where the excitation energies $ \epsilon_i $ are disordered.
The initial excitation at site 1 is transferred
to the terminal site $L+1$
and subsequently trapped there in a reservoir with rate $\kappa$.
The excitation also decays at each site with rate $\Gamma$.
(The effects of $\kappa$ and $\Gamma$ will be included properly later.)
It is well known that the excitation is localized in this sort
of disordered system (Anderson localization) \cite{Localization}.
By using a perturbation theory \cite{Kato}
and neglecting the rapidly oscillating terms,
the average population (time-averaged pseudosteady state) at the site $L+1$ is calculated in the leading order as
%%%
\begin{eqnarray}
\bar{p}_{L+1} \sim (L+1)(v/\epsilon)^{2L}
\end{eqnarray}
%%%
for $ v / \epsilon \ll 1 $, where $\epsilon \sim |\epsilon_i - \epsilon_j|$
($i\not=j$) denotes the degree of disorder.  This localization certainly
suppresses the excitation transfer to the reservoir.
%%%
%%%

It has been discussed that by repeatedly measuring the location of excitation
the delocalization may occur as the anti-Zeno effect \cite{Delocalization}.
Here, we identify properly the crossover from the Zeno to the anti-Zeno regimes
with respect to the energy transfer
by determining the time evolution of the system under repeated measurements.
The density matrix after the $n$ (non-selective) measurements
of the excitation with an interval $\tau$ becomes 
%%%
\begin{eqnarray}
\rho (t_n) = \sum_{i=1}^{L+1} p_i (t_n) | i \rangle \langle i | ,
\label{RepeatedMeasurement}
\end{eqnarray}
%%%
where the probability to find the excitation at site $ i $
for $ t_n = n \tau $ is given by
%%%
\begin{eqnarray}
p_i (t_n) = \langle i |
e^{- i H \tau} \rho (t_{n-1}) e^{i H \tau} | i \rangle .
\end{eqnarray}
%%%
In particular, with a small enough $\tau < 1 / \epsilon \ll 1/v $
(frequent measurements) under the large disorder $ \epsilon \gg v $,
we obtain the recursive equation
$ p_i (t_n) \approx (1- 2 \tau^2 v^2) p_i (t_{n-1})
+ \tau^2 v^2 [p_{i-1}(t_{n-1})+ p_{i+1}(t_{n-1})]$.
Then, for $ n \geq L $ and $ n (\tau^2 v^2) < 1 $ ($ \tau v < 1/{\sqrt L} $)
the population at site $L+1$ amounts approximately to
%%%
\begin{eqnarray}
p_{L+1}(t_n) \approx \left( \begin{array}{c} n \\ n-L \end{array} \right)
( 1 - 2 \tau^2 v^2)^{n - L} (\tau^2 v^2)^L .
\end{eqnarray}
%%%
This means that the excitation is transferred almost straightforwardly
with little backward transfer.

We can identify the Zeno and anti-Zeno regimes for site $L+1$
by the conditions $ p_{L+1}(t_n) < \bar{p}_{L+1} $
and $ p_{L+1}(t_n) > \bar{p}_{L+1} $, respectively,
in comparison with the no measurement case,
following the treatment in Ref. \cite{Luis03}.
Since $ p_{L+1}(t_n) \sim ( n^L / L^L ) (\tau^2 v^2)^{L} $
for $ n ( \tau^2 v^2 ) < 1 $ by considering roughly the behavior
of the binomial coefficient, the crossover time is found as
$ t_{\rm c} = n_{\rm c} \tau \sim L /(\epsilon^2 \tau) $.
The condition $ n_{\rm c} \geq L $ implies $ \tau \lesssim 1/\epsilon $.
This result clearly indicates that for the large disorder $ \epsilon$
the Zeno regime rapidly changes to the anti-Zeno regime
with the small $ t_{\rm c} $.
It is also found that in the one-dimensional system the time $ t_{\rm c} $
needed for the propagation of anti-Zeno effect to the trapping site
is roughly proportional to the length $ L $ of the system.

%%%%%
{\it Efficiency of energy transfer.}---
%%%%%
To obtain close insights, we first examine the two-site case
($L+1=2$), where the relevant calculations can be made analytically.
In general, the fully connected multisite systems
have direct coupling between the initially excited site
and the trapping site.
Thus, it is expected that the results for the two-site model
concerning the energy transfer under the large disorder
are essentially valid even for the fully connected systems.
On the other hand, if the initial site does not have direct coupling
to the trapping site, the excitation would be transferred through
some intermediate sites along the shortest path to the trapping site.
In such a case, however, the energy transfer would be diminished
substantially, since it takes more time to propagate the anti-Zeno effect
to the trapping site.
These features will be confirmed by numerical calculations,
as discussed in detail later.

In the two-site model, the initial excitation at site 1 is transferred
to site 2 and then trapped in the reservoir with rate $\kappa$.
The excitation is also dissipated at each site
with rate $\Gamma$ by the interaction with the environment.
These effects are taken into account in the Hamiltonian by replacing
$\epsilon_1$ and $\epsilon_2$ in Eq. (\ref{Anderson}) for $L+1=2$
with $\epsilon_1-i\Gamma$ and $\epsilon_2-i\kappa-i\Gamma$, respectively.
The efficiency of energy transfer is given by
%%%
\begin{eqnarray}
\eta (\tau) = 2 \kappa \int _{0} ^{\infty} dt
\langle 2 | \rho (t) | 2 \rangle
= 2 \kappa \int_{0}^{\infty} dt p_{2}(t) .
\end{eqnarray}
%%%
It is calculated by using the analytic expression of $ p_{i}(t) $
with the $ 2 \times 2 $ transition probability matrix
for sites 1 and 2 which is determined
by the spectral decomposition of the time evolution operator $ e^{-iHt} $.

\begin{figure}
\centering
\scalebox{0.9}
{\includegraphics*[1.5cm,1.7cm][9.5cm,6.3cm]{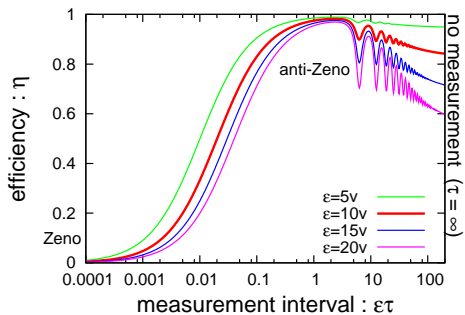}}
\caption{(Color online) Efficiency $\eta(\tau)$ as functions of the measurement interval
$\tau$ (normalized with $ \epsilon $).  We have taken typically
$ \epsilon \equiv \epsilon_1 - \epsilon_2 = 5v,10v,15v,20v$
(from top to bottom), $\kappa =0.5v$ and $\Gamma = 0.001v$.}
\label{Efficiency}
\end{figure}

The results for $ \eta (\tau) $ are plotted in Fig. \ref{Efficiency}
as functions of the measurement interval $\tau$
for the various degrees of disorder (energy gap)
$\epsilon \equiv \epsilon_1 - \epsilon_2 = 5v, 10 v, 15v, 20v$
(from top to bottom) with the typical $\kappa=0.5v$ and $\Gamma=0.001v$.
In the case of $\tau \to 0$,
the efficiency converges to zero due to the quantum Zeno effect.
On the other hand, the maximal efficiency is achieved
for $ \tau \simeq \pi/\epsilon $
under the large disorder $ \epsilon \gg v \sim \kappa $ as
%%%
\begin{eqnarray}
\eta ( \tau \simeq \pi/\epsilon  ) \simeq
1 - 2 \Gamma \left (\frac{1}{\kappa} + \frac{\pi \epsilon}{4 v^2} \right)
= 1 - O ( \Gamma \epsilon / v^2 ) ,
\label{eta-max}
\end{eqnarray}
%%%
where the hopping rate of excitation from site 1 to site 2
for the measurement interval $ \tau $ becomes maximal approximately
as $ 4 ( v / \epsilon )^2 $.
In fact, it is seen in Fig. \ref{Efficiency}
that the efficiency $ \eta ( \tau ) $ is enhanced optimally
via the anti-Zeno effect, approaching unity around $ \epsilon \tau \sim 1 $,
where the disorder is actually compensated by the energy uncertainty
due to the measurement.
The deterioration of the efficiency is only linear
with respect to the disorder $ \epsilon $ in Eq. (\ref{eta-max}).
This should be compared with the deterioration
for the case without measurements as
%%%
\begin{eqnarray}
1 - \eta (\tau \to \infty) \sim (\Gamma/\kappa) (\epsilon/v)^2 ,
\label{eta-nomeasurement}
\end{eqnarray}
%%%
growing quadratically with the large disorder $ \epsilon \gg v $.
In this case, the efficiency diminishes substantially
with suppression of the effective trapping rate
as $ \kappa {\bar p}_2 \sim \kappa (v/\epsilon)^2 $
due to the localization.
Here, it should be noted, as seen in Eq. (\ref{eta-max}),
that an essential condition for the high efficiency
$ \eta ( \tau \sim 1/\epsilon ) $ is to suppress the dissipation sufficiently
as $ \Gamma / v \ll v/\epsilon \ll 1 $ ($ \kappa \sim v $).
Specifically, $ \eta ( \tau \sim 1/\epsilon ) \approx 98 \% $
and $ \eta (\tau \to \infty) \approx 80 \% $
for $ \epsilon = 10v $, $ \kappa = 0.5v $ and $ \Gamma = 0.001v $,
as seen in Fig. \ref{Efficiency}.
%%%
%%%

In order to obtain further understanding on the energy transfer,
we have also made numerical calculations for the multisite cases
including the one-dimensional models (two to ten sites) in Eq. (\ref{Anderson})
and fully connected models (three and four sites)
with coupling for each pair of sites as
$ v_{ij} |i \rangle \langle j| + v_{ji} |j \rangle \langle i| $.
The site energies $ \epsilon_i $ and couplings $ v_{ij} = v_{ji} $
(real for simplicity) are taken randomly
as $|\epsilon_i - \epsilon_j| \sim \epsilon $
and $ v_{ij} \sim v $ ($ i \not= j $),
with $ \epsilon = \epsilon_1 - \epsilon_{L+1} $ as the mean disorder
and $ v = v_{1L+1} $
between the initial site 1 and the trapping site $ L+1 $.
It is first found in any model that the optimal condition
on the measurement time interval for the anti-Zeno effect
is given by $\tau \sim 1/ \epsilon $,
as seen in Eq. (\ref{eta-max}) for the two-site model.
This result implies that the excitation is transferred efficiently
by virtue of the energy-time uncertainty 
introduced by the repeated measurements.
It is also confirmed that the fully connected models exhibit
essentially the same behavior for the efficiency $ \eta ( \tau ) $
as the two-site model under the large disorder $ \epsilon \sim 10v $.
On the other hand, in the one-dimensional $(L+1)$-site models
($ L+1 \geq 3 $) with the intermediate sites, the maximal efficiency
$ \eta ( \tau \sim 1 / \epsilon ) $ deteriorates by an amount roughly
$ L ( \Gamma \epsilon / v^2 ) $ due to the net loss
through the sites with the length $ L $ of the system.
The efficiency for the case of no measurement
diminishes substantially as $ \eta ( \tau \to \infty ) \approx 0 $
due to the highly suppressed trapping rate
$ \kappa {\bar p}_{L+1} \sim \kappa (L+1) (v/\epsilon)^{2L} $
with $ \epsilon \sim 10v $ under the localization.
It should further be remarked that the specific four-site model,
where only the coupling between sites 1 and 4 is cut off
($ v_{14}=0 $), behaves like the one-dimensional three-site model
with somewhat higher efficiency achieved
through the two $ L=2 $ paths to trapping site 4.
These results indicate that the initial excitation is transferred
along the shortest path.

%%%%%
{\it Decoherence induced enhancement.}---
%%%%%
Here, we note that environment acts as frequent measurements on the system.
Specifically, the master equation with dephasing under the environment
can be unraveled as a stochastic quantum jump,
%%%
\begin{eqnarray}
\rho (t + \Delta t)
&=& (1- 2 \gamma \Delta t) e^{-i H \Delta t}\rho (t)e^{iH\Delta t}
\nonumber \\
&{}& + 2 \gamma \Delta t \sum_i  P_i \rho(t) P_i ,
\end{eqnarray}
%%%
where $ P_i \equiv | i \rangle \langle i | $,
and $ \gamma $ is the dephasing rate.
Then, it reproduces the time evolution
of the density matrix in Eq. (\ref{RepeatedMeasurement})
under a relevant assumption that the jumps occur periodically
in the small interval $\tau = 1/(2 \gamma)$ \cite{Jump}.
Thus, the results obtained so far in the model systems
under repeated measurements are applicable
for the environment-assisted quantum transport,
which has been revealed recently for light-harvesting photosynthetic
complexes \cite{Rebentrost09,Plenio08}.  In fact, the two-site model
(Fig. \ref{Efficiency}) reproduces essentially the result obtained
in a coupled seven-site system for the FMO complex
with the direct coupling between the initial and trapping sites
(Fig. 2 in Ref. \cite{Rebentrost09}).
The parameters adopted in Fig. \ref{Efficiency} are actually comparable
to those estimated for the FMO complex in Ref \cite{Adolphs06}.
Therefore, the decoherence-induced enhancement of energy transfer
in disordered systems may be understood as the quantum anti-Zeno effect.
The result in Eq. (\ref{eta-max}) for the two-site model
suggests that even in the presence of significant disorder
$ \epsilon \sim 10v $ the quite high efficiency in the FMO complex
is achieved under the optimal condition
$ 2\gamma \sim \epsilon $ (i.e., $ \tau \sim 1/\epsilon $)
between the dephasing and disorder with the small enough dissipation
$ \Gamma \sim 10^{-3} v $ ($ \kappa \sim v $).
We also note that the deterioration of efficiency $ \sim 20 \% $
for the case with negligible dephasing (no measurement),
as shown in Ref. \cite{Rebentrost09},
is in reasonable agreement with Eq. (\ref{eta-nomeasurement})
for the two-site model, where the effective trapping rate is suppressed
as $ \kappa {\bar p}_2 \sim \kappa (v/\epsilon)^2 $ due to the localization.
In these observations, the simple two-site model seems to describe
relevantly the energy transfer in the FMO complex.
This will be because the direct coupling is present
between the antenna pigment and the trapping site, as shown numerically
for the fully connected multi-site (three and four sites) models.

\begin{figure}
\scalebox{0.85}
{\includegraphics*[1.8cm,1.8cm][9.5cm,6.5cm]{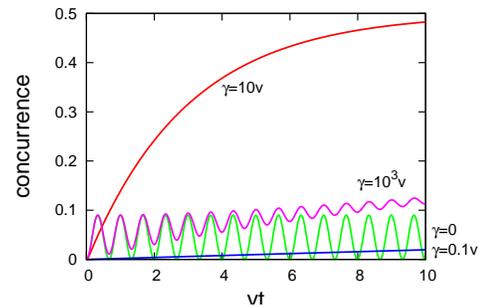}}
\caption{(Color online) 
Concurrence as functions of time $t$ (normalized with $v$)
for excitation energies $ \epsilon_1 = \epsilon_3 = v $ and
$ \epsilon_2 = 10v $ and dephasing rates
$2 \gamma = 0, 0.1v, 10v, 10^3 v$.}
\label{Entangle}
\end{figure}

%%%%%
{\it Promotion of entanglement generation.}---
%%%%%
We show another interesting phenomenon related to the anti-Zeno effect.
Although it requires a specific condition,
which would not be met in general disordered systems,
not only the population but also truly quantum quantities
such as coherence or entanglement can be enhanced
by the repeated measurements (or decoherence).
Consider a linearly coupled three-qubit system
in Eq. (\ref{Anderson}) with $L+1=3$,
where the excitation energies at sites 1 and 3
are specifically degenerate as $ \epsilon_1 = \epsilon_3 $.
The initial excitation located at site 2
is transferred to neighboring sites 1 and 3.
Then, some amount of entanglement is generated
between sites 1 and 3 even if site 2 is traced out.
The concurrence as a measure of entanglement
between sites 1 and 3 \cite{Wootters98}
is calculated for the case without measurements as
%%%
\begin{eqnarray}
C(t) =  \frac{4v^2}{8 v^2+ \epsilon^2}
\left[ 1 - \cos \left( \sqrt{8 v^2+ \epsilon ^2}t \right ) \right] ,
\end{eqnarray}
%%%
where $\epsilon = \epsilon_2 - \epsilon_1 = \epsilon_2 - \epsilon_3$.
This shows that the generated entanglement is very small
for the large disorder $ \epsilon \gg v $ due to the localization.
On the other hand, in the case where the qubit at site 2
is measured with the interval $\tau$,
the concurrence $ C_\tau (t) $ is given in terms of $C(t=\tau)$ as
%%%
\begin{eqnarray}
C_\tau (t) = \frac{1}{2} \left\{ 1 - [ 1 - 2 C(\tau)] ^{t/\tau} \right\} .
\label{concurrence}
\end{eqnarray}
%%%
Note here that $ C_\tau (\tau) = C(\tau) $
since the measurement is not made until $ t=\tau $.
The same as the energy transfer, in the limit of $\tau \to 0$
for $ C(\tau) \to 0 $ with finite $t$,
the concurrence under the repeated measurements
does not grow due to the Zeno effect.
On the other hand, in the limit of $t \to \infty$
with finite $\tau$ for $ 2 C(\tau) < 1 $,
it converges to 1/2 far beyond the value without measurements,
even in the presence of large disorder $\epsilon \sim 10v$.
These results are also valid
when the system is exposed to dephasing due to the environment.
Actually, we have solved numerically the master equation
to determine the time evolution of the system.
We plot in Fig. \ref{Entangle} the resultant concurrence
as functions of time $t$
for excitation energies $ \epsilon_1 = \epsilon_3 = v $ and
$ \epsilon_2 = 10v $ and dephasing rates $2 \gamma =0, 0.1v,10v, 10^3v$.
These results agree with $C_{\tau}(t)$ in Eq. (\ref{concurrence})
by setting $\tau = 1/(2 \gamma)$.
Thus, we find in this case that the local decoherence
($ 2 \gamma \sim \epsilon \sim 10v $) promotes
the entanglement generation between the nonlocal sites.
%%%
%%%

Although some processes for entanglement generation
through correlated environments or selective measurements
have been discussed so far \cite{Braun02,Nakazato04},
they are certainly different from the present case
where the local decoherence or non-selective measurements
promote the generation of non-local entanglement.
There are also stochastic resonance-like phenomena,
where entangled steady states are achieved for an optimal decoherence rate
\cite{Hartmann06,Huelga07,Wittaut08}.
In these cases, decoherence serves as an initialization (reset) mechanism.
The present case, on the other hand, can be understood
by considering the fact that the excited states at sites 1 and 3
span a sort of Zeno subspace \cite{Facchi02}
as $ | 1 \rangle \langle 1 | + | 3 \rangle \langle 3 |
= I - | 2 \rangle \langle 2 | $,
which is not disturbed by the measurement of site 2.
The degeneracy $ \epsilon_1 = \epsilon_3 $ between sites 1 and 3
is an essential condition to promote significantly the entanglement
generation. In this case (sites 1 and 3 couple to site 2
with the same strength $v$), the specific combination
$(| 1 \rangle + | 3 \rangle)/{\sqrt 2}$ couples coherently
with $ | 2 \rangle$ in the time evolution,
while the orthogonal one $(| 1 \rangle - | 3 \rangle)/{\sqrt 2}$ decouples.
Then, the transfer of the initial excitation to the Zeno subspace
is accelerated by repeated measurements or dephasing.

%%%%%
{\it Discussion.}---
%%%%%
We have investigated how repeated measurements
or dephasing by the environment affect quantum transport
in disordered systems.  Specifically, the quantum anti-Zeno effect
can enhance the processes such as energy transfer
and entanglement generation.
The two-site model (or multisite model with direct coupling
between the initial and trapping sites) reproduces essentially
the numerical calculation for the photosynthetic FMO complex,
suggesting that the energy transfer in biological systems is enhanced
by the quantum anti-Zeno effect.

Despite thier simplicity, the multi-site models,
especially the two-site case, which we have investigated here,
have good potential to reproduce various important phenomena
including energy transfer.
For example, as seen in Fig. \ref{Efficiency},
in the regime ($ \tau \to \infty $) of no measurement
or negligible dephasing the population at trapping site 2 is very
sensitive to the disorder (energy gap) due to the Anderson localization.
Thus, if the disorder (energy gap) originates from the external field
(e.g., Zeeman shift), the yield of a certain chemical product,
which is synthesized in the reserver coupled to site 2,
may serve as a sensor of the external field.
This kind of chemical compass has been thought of
as the mechanism underlying the magnetic sensitivity
of certain migratory birds \cite{Maeda08,Kominis08,Cai09,Gauger09}.
Actually, the two-site model is qualitatively consistent
with the singlet-triplet production in the radical pair model,
including the fact that the sensitivity is degraded by decoherence
\cite{Kominis08,Cai09,Gauger09}.
On the other hand, in the transition between the Zeno and the anti-Zeno regimes
the efficiency of energy transfer changes very sensitively
with the dephasing rate $\gamma = 1/(2 \tau)$ depending on the temperature.
Thus, this provides a temperature sensor,
which might be functioning in biological systems.

As seen so far, simple quantum systems such as the two-site model
may be used as starting points to understand various mechanisms
and properties of chemical and biological systems,
where both disorder and decoherence are certainly in existence.
We believe that deeper understanding of the sophisticated mechanisms
in natural systems will present us with new ways for robust quantum control
in noisy physical devices.

%%%%%
\begin{acknowledgments}
The authors would like to thank Takahiro Sagawa and Toru Kawakubo
for valuable comments.
This work was supported by the JSPS Grant No. 20$\cdot$2157.
\end{acknowledgments}
%%%%%

\end{document}